\documentclass[10pt,conference]{IEEEtran}
\IEEEoverridecommandlockouts
\usepackage{cite}
\usepackage{amsmath,amssymb,amsfonts}
\usepackage{algorithmic}
\usepackage{graphicx}
\usepackage{textcomp}
\usepackage{xcolor}
\usepackage{url}
\def\BibTeX{{\rm B\kern-.05em{\sc i\kern-.025em b}\kern-.08em
    T\kern-.1667em\lower.7ex\hbox{E}\kern-.125emX}}

\usepackage{siunitx}
\usepackage[group-separator={,}]{siunitx}

\begin{document}

\title{LogLead -- Fast and Integrated Log Loader, Enhancer, and Anomaly Detector
\thanks{Funded by Research Council of Finland grant id 349487}
}
\author{\IEEEauthorblockN{1\textsuperscript{st} Mika V. Mäntylä}
\IEEEauthorblockA{\textit{Department of Computer Science} \\
\textit{University of Helsinki}\\
Finland \\
mika.mantyla@helsinki.fi}
\and
\IEEEauthorblockN{2\textsuperscript{nd} Yuqing Wang}
\IEEEauthorblockA{\textit{Department of Computer Science} \\
\textit{University of Helsinki}\\
Finland \\
yuqing.wang@helsinki.fi}
\and
\IEEEauthorblockN{3\textsuperscript{rd} Jesse Nyyssölä}
\IEEEauthorblockA{\textit{Department of Computer Science} \\
\textit{University of Helsinki}\\
Finland \\
jesse.nyyssola@helsinki.fi}

}
\maketitle

\begin{abstract}
This paper introduces LogLead, a tool designed for efficient log analysis benchmarking. LogLead combines three essential steps in log processing: loading, enhancing, and anomaly detection. The tool leverages Polars, a high-speed DataFrame library. We currently have Loaders for eight systems that are publicly available (HDFS, Hadoop, BGL, Thunderbird, Spirit, Liberty, TrainTicket, and GC Webshop). We have multiple enhancers with three parsers (Drain, Spell, LenMa), Bert embedding creation and other log representation techniques like bag-of-words. LogLead integrates to five supervised and four unsupervised machine learning algorithms for anomaly detection from SKLearn. By integrating diverse datasets, log representation methods and anomaly detectors, LogLead facilitates comprehensive benchmarking in log analysis research. We show that log loading from raw file to dataframe is over 10x faster with LogLead compared to past solutions. We demonstrate roughly 2x improvement in Drain parsing speed by off-loading log message normalization to LogLead. Our brief benchmarking on HDFS indicates that log representations extending beyond the bag-of-words approach offer limited additional benefits. Tool URL: \url{https://github.com/EvoTestOps/LogLead}







\end{abstract}


\section{Introduction}
Due to the rising popularity of DevOps in software development and the need to avoid runtime failures, log analysis has gained importance. Log analysis not only enhances runtime quality but also aids in software testing, root-cause analysis, and security investigations. Research tooling in this area seems insufficient. To the best of our knowledge, and based on a 2021 literature review \cite{lupton2021literature}, Loglizer \cite{he2016experience} is currently the only toolkit available for log anomaly benchmarking. Another notable research tool is Aminer by Landauer et al. \cite{landauer2019framework}, but its focus is on runtime security analysis, not benchmarking. Industry tools on log analysis, such as Splunk, Microsoft Sentinel, and Elastic\footnote{AKA Security Information and Event Management (SIEM) tools}, are abundant. However, they have limited free options for incorporating custom algorithms or for benchmarking, making them ill-suited for research.

In this paper, we present LogLead\footnote{\url{https://github.com/EvoTestOps/LogLead}} (\underline{Log} \underline{L}oader, \underline{E}nhancer, \underline{A}nomaly \underline{D}etector), a free open-source tool for log analysis benchmarking. The tool has two goals motivating our work.

First, we want to accelerate research in log anomaly detection by offering an integrated log analysis benchmarking tool. LogLead should enable researchers to spend more effort on innovation and less on the somewhat mundane task of benchmarking their work against previous studies. Currently, it features public log data from 8 systems, 7 log enhancement methods, and 11 machine learning classifiers, seamlessly integrated for end-to-end functionality. This integration offers over 600 unique combinations (8 × 7  × 11), against which researchers can compare their work.


The second goal is to enhance the computational efficiency of log analysis. As software logs can be be generated at massive volume up to PBs per day \cite{li2023logshrink}, 
it's vital that all operations performed on the log files are fast and efficient.

The upcoming sections detail the tool's overall architecture and its main components in Sections II-V. We also offer a brief empirical evaluation, and compare our tool to prior work. Section VI discusses current limitations and future development ideas, while Section VII presents the conclusions.










\section{Overview}
\subsection{Technology and Architecture}
The performance of any tool depends on the technologies it is built upon. Therefore, we chose to build our tool on the Polars library \cite{Vink2023Polars}, which is advertised as a ``blazingly fast DataFrame library". Polars is built on the Rust implementation of the Apache Arrow\footnote{https://github.com/jorgecarleitao/arrow2} specification. Apache Arrow aims to enhance data analytics performance and standardize data representation in memory.

\begin{figure}[ht]
\centering
\includegraphics[width=0.42\textwidth]{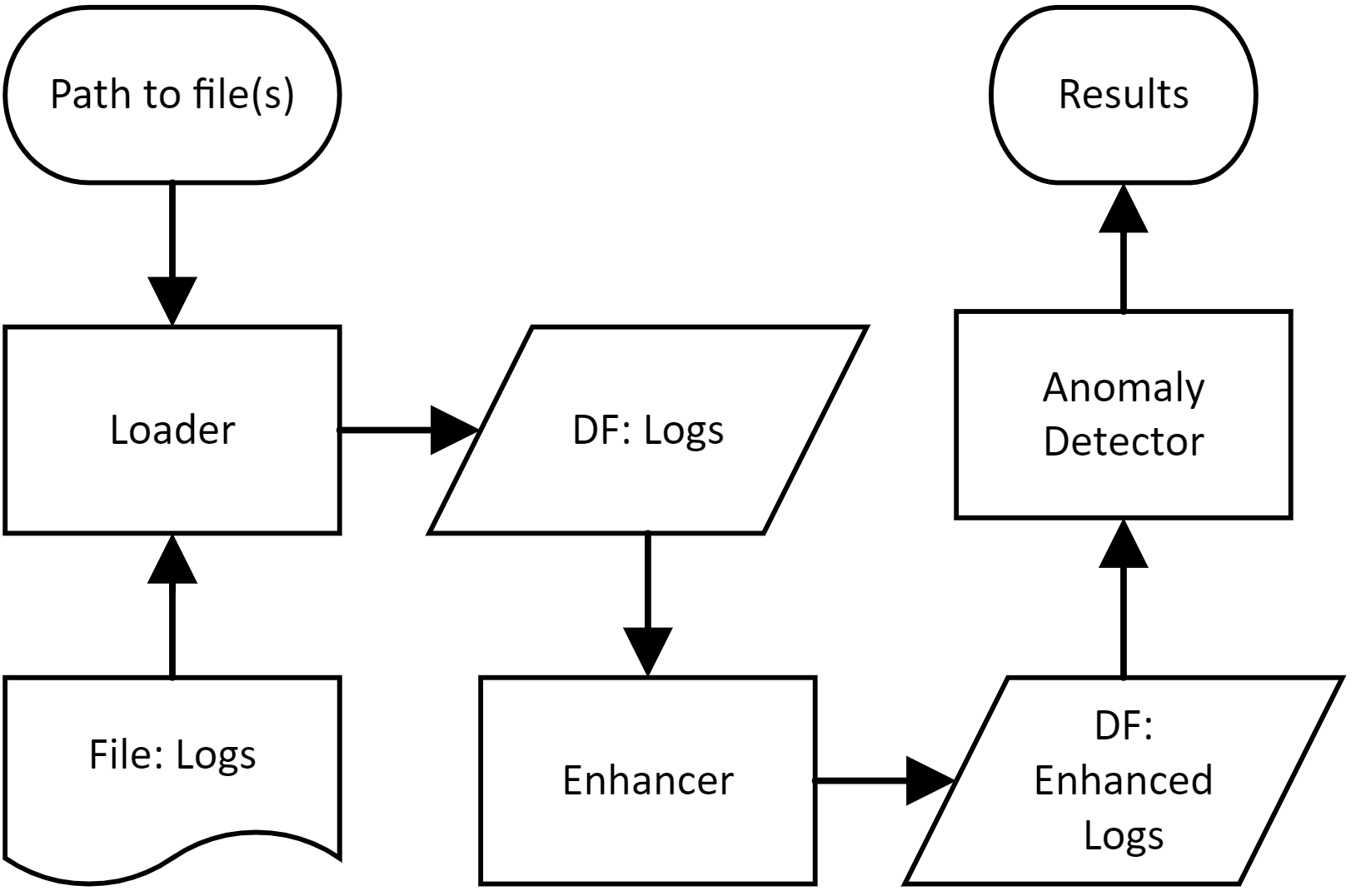}
\caption{LogLead Dataflow Diagram}
\label{fig:arch}
\end{figure}

The architecture of LogLead comprises three components, as shown in Figure \ref{fig:arch}. The Loader is responsible for loading log files into dataframes and performing log-file-specific pre-processing. The Enhancer enhances the dataframes by creating various representations of logs. Finally, the Anomaly Detector uses the enhanced log data to perform anomaly detection. 

Prior research in log analytics has used Pandas \cite{he2016experience, zhu2019tools}. No academic comparisons of Polars and Pandas exist, but grey literature suggests that Polars significantly outperforms Pandas,  with statements like ``Polars is really fast... Polars wipes the floor with it [Pandas]" \cite{Heavey2023Modern}.

We compared our tool to Loglizer tool \cite{he2016experience}, by examining the LogLizer GitHub repository\footnote{https://github.com/logpai/loglizer}, which was updated as recently as 2023. 
Loglizer supports two public datasets (HDFS and BGL), compared to LogLead's support for 8 public and one private dataset. Notably, Loglizer assumes parsed data, while LogLead handles raw data. LogLead also offers more log enhancement options than Loglizer. In anomaly detection Loglizer relies on parsed log messages for predictions, whereas LogLead can utilize multiple log representations.

\subsection{Tool Usage}

LogLead is used through Python code or in a notebook-style environment, where individual cells can be executed separately or combined into larger scripts. We decided against developing a separate user interface, as researchers and practitioners in log analysis are typically familiar with notebooks or Python. This approach also enables users to easily supplement any missing features directly on top of Polars dataframes. To aid easy onboarding, there are two demo files in LogLead's github\footnote{\url{https://github.com/EvoTestOps/LogLead/}}.
\section{Loader}
\subsection{Input-Output}
Loader's input includes paths to log files and, if present, label files. It outputs dataframes with standardized names. A single dataframe is produced, each line holding a log event, if labels are not present or if log anomalies are labeled at the line (event) level. When labels are provided at the sequence level, two dataframes on different levels are created: the lower (Event) level dataframe, where each row is a log event, and a higher (Sequence) level dataframe, where each line holds a sequence ID and the status of the sequence (normal/anomaly). In SQL terminology, the `sequence ID' serves as the primary key in the sequence dataframe, establishing a link to the event dataframe, where the same `sequence ID' is used as a foreign key. More data is later aggregated to the sequence-level dataframe. The approach for creating Sequence dataframes in LogLead differs depending on the log data. For instance, in HDFS \cite{xu2009detecting}, sequences are created using the block\_id present in log message of each event. In contrast, for systems like Hadoop \cite{lin2016log} and Android stability testing (private data \cite{mantyla2022pinpointing}), sequences are derived from file and folder names.

\subsection{Architectural Responsibility}
Software log files vary across systems. Variations include the fields present in the logs; the representation of timestamps; the file format and structure of the logs, e.g., single-file, multi-file, multi-folder; and how anomalies, if any, are labeled within the logs. Additionally, raw log files often contain peculiarities such as incomplete log lines, non-UTF characters, and stack traces that disrupt the continuous flow of log events.

Loader's primary responsibility is to hide the log-specific differences and peculiarities from other components. It also ensures that once a system's log files are loaded, the resulting dataframe appears as standardized as possible. Achieving completely identical dataframes is not feasible due to the varying amounts of information in different logs. Additionally, it is crucial not to lose any information from the logs, as it might be important for analysis. Thus, LogLead enforces standardized names for columns holding the same data across different logs. 

\subsection{Detailed description}
Loader is implemented as a Python module, with the BaseLoader class encapsulating shared functionalities. Classes extending BaseLoader are designed to handle the peculiarities of each system's log files.

Each Loader takes the location of log files as input and loads the files. It splits the data from the logs into the correct columns within a dataframe. Additionally, it converts timestamps from strings to Polars Time objects. Loader manages the labels (anomaly/normal), which may be part of each log line, included in a separate file, or indicated by the file name. If log labels are given at the sequence level rather than per log event (line), it also generates a sequence-level dataframe, where each sequence is represented once.

BaseLoader class includes functionality to check each Loader's dataframe for null values, issuing a warning if any are found, as null values can cause issues in subsequent components. 
Additionally, BaseLoader verifies the consistency of naming across all log files as every log line should have a log message and a timestamp, which are named ``m\_message" and ``m\_timestamp" respectively in the dataframe.

We currently have loaders implemented for nine systems, representing a range of diverse log data sources. These include four supercomputers: BGL, Thunderbird, Spirit, and Liberty \cite{oliner2007supercomputers}; two well-known public datasets of HDFS \cite{xu2009detecting}, and Hadoop \cite{lin2016log}. These six datasets are all easily available\footnote{\url{https://github.com/logpai/loghub} or \url{https://www.usenix.org/cfdr-data}}. We also have logs from two microservice systems: TrainTicket and the Google Cloud Web Shop demo from the Nezha paper~\cite{yu2023nezha}\footnote{\url{https://github.com/IntelligentDDS/Nezha}}. There's also a private dataset loader for Android data \cite{mantyla2022pinpointing}.
We have a loader for the Graylog Extended Log Format (GELF), though we currently lack a dataset for it. Lastly, our general-purpose loader, which does not split data into columns, is useful for evaluating whether LogLead is suitable for processing an arbitrary log file.

\subsection{Empirical Evaluation}
The code for all results is in LogLead's GitHub\footnote{\url{https://github.com/EvoTestOps/LogLead/tree/main/demo/paper}}. All tests were executed on a VM with 28 Intel CPUs (Broadwell, IBRS) at 2.4 GHz, 224 GB RAM, and two NVIDIA Tesla P100 PCIe 16 GB GPUs. LogLead also works on laptops and desktops, if RAM is sufficient for the log file.

To assess the performance efficiency of our Loader component, we study the fundamental task of loading logs from their original raw text file into a dataframe. This step is essential and must be completed before any log analysis can begin. We compared this process with the equivalent steps employed in the Logparser project\footnote{\url{https://github.com/logpai/logparser/}} \cite{zhu2019tools}, which utilizes Pandas dataframes.

From Table \ref{tab:load}, it is evident that LogLead significantly outperforms the Logparser in terms of loading speed. Specifically, LogLead loads log files to memory 14.8 and 15.1 times faster than Logparser for HDFS and BGL datasets, respectively. In other words, LogLead's loading process takes 6.8\% and 6.6\% of the time required by the Logparser for the HDFS and BGL datasets. Unfortunately, loading times for Thunderbird and Hadoop logs are not available for comparison with Logparser. The Hadoop logs, which are spread across multiple files and folders, present a challenge for the Logparser loader, as it is designed to load data from single files only. As for the Thunderbird logs, despite several attempts, the Logparser loader failed to return a complete dataframe, consistently stalling after processing approximately 34 million lines.

Table \ref{tab:load} demonstrates that LogLead's load time scales linearly, achieving an approximate speed of processing roughly one million rows per second in a single file case, as observed with BGL, HDFS, and Thunderbird datasets. However, the more complex multi-file and multi-folder structure of log files, such as with the Hadoop data, results in proportionally slower loading times. While our data loading from multiple files is parallelized, it appears that the Hadoop data's structure might be too small to fully benefit from this parallelization.

\begin{table}
\centering
\caption{Loading times in seconds from file to dataframe}
\label{tab:load}
\begin{tabular}{l r S S}   
{Data} &\multicolumn{1}{l}{Log Lines}&{LogLead (Polars)}&{Logparser \cite{zhu2019tools} (Pandas)}\\
\hline 
Hadoop & 177,592 & 2.31 & NA \\ 
BGL &4,747,963& 5.40 &  81.55\\
HDFS &11,175,629& 9.16 & 135.74 \\
Thunderbird &211,212,192& 269.28 & NA  \\
\end{tabular}
\end{table}

\subsection{Comparison to related work}
For loading Logparser \cite{zhu2019tools} uses a single line log format specification, which is then converted into a regular expression. This method is  elegant as it eliminates the need for log-specific loaders. However, it's unclear whether this approach can handle incomplete log lines or stack traces. A drawback of their method is its performance. This reduced efficiency is due to the use of a Python loop for splitting each line into columns, whereas we leverage native Polars operations. Additionally, Polars currently lacks support for splitting a line using a regular expression, precluding us from adopting the more elegant approach of the Logparser project. Therefore, there's a notable tradeoff between two quality attributes here. LogLead excels in performance efficiency, while the Logparser project has an advantage in terms of code maintainability.

\section{Enhancer}
\subsection{Input-Output}
Enhancer receives the dataframe(s) from Loader. It then applies the enhancements to these dataframe(s). The output of Enhancer is the enhanced dataframe(s), see Figure \ref{fig:arch}.

\subsection{Architectural Responsibility}
Enhancer is tasked with log file independent log enhancements, also known in past work as log representation \cite{wu2023effectiveness}. Another term we could use is feature extraction, derived from machine learning. Since certain parts of logs, like log messages, are in natural language, many enhancement techniques stem from NLP, including bag-of-words and word embeddings. Log enhancement techniques that focus particularly on reducing the feature space by grouping log messages originating from the same logging call in the source code are known as log parsing, log message template extraction, log key extraction, or log message clustering \cite{zhu2019tools}. Enhancer encapsulates these log-file independent enhancements and applies them to dataframes where the log files are loaded.

\subsection{Functional Description}
At the event level, we offer multiple log enhancements such as normalizing log messages, converting log messages into word lists, parsing with Drain \cite{he2017drain}, Spell \cite{du2016spell}, and LenMa~\cite{shima2016length}, and transforming log messages into BERT embeddings \cite{devlin2018bert}, also referred to as Neural Parser \cite{le2021log}. Log message normalization involves simplifying log messages, for instance, replacing all IP addresses in a log message with a fixed token \textless IP\textgreater. LogLead allows users to provide a custom regular expression for normalizing log messages. LogLead utilizes Polars for executing the Regex, which depends on Crate\footnote{https://docs.rs/regex/latest/regex/}. This means that regular expressions are executed quickly, but advanced features like look-ahead and look-behind are not available.

Turning log messages into a list of words enables the later Anomaly Detector component to use a bag-of-words-based anomaly detection method. Log parsing with Drain, Spell, or LenMa converts each log message into an event ID. 
Log message event ID streams can be employed for next event prediction using the n-gram model we have integrated from \cite{mantyla2022pinpointing}, or they can be directly used in the Anomaly Detector.
Additionally, we support transforming log messages into sentence embeddings using either the BERT base model
or the ALBERT base model.
Both models have 12 layers of transformer encoder and 768 hidden units of each transformer. We use the word embeddings generated by the last encoder layer of the model, and calculate the sentence embedding of a log message as the average of its word embeddings.

On the sequence level, we aggregate information from individual log events within sequences. Past research, such as that by Landauer et al. \cite{landauer2023deep}, has demonstrated the utility of simple aggregated sequence-level metrics. For instance, they found that in the HDFS dataset, any sequence shorter than the shortest normal sequence is an anomaly. In response to this finding, our tool includes the capability to aggregate various such measures into a sequence-level dataframe. For example, we can measure sequence length in terms of the number of events or the duration of the sequence. Additionally, it is possible to aggregate log event-level representations, like parsed log message IDs or words, into a list of event IDs or words at the sequence level.

\subsection{Empirical evaluation}
Log parsing is a widely-used method for creating new representations of log messages, with Drain \cite{he2017drain} being one of the most efficient and popular parsers. Log normalization, referred to as masking in Drain, can be performed either in LogLead or directly in Drain configured via the Drain.ini file. Both LogLead and Drain support the use of regular expressions to generate a normalized form of the log messages.

We conducted tests on the parsing speed with our integration of Drain under two different conditions. In the first scenario, Drain was used for both log file normalization and parsing. In the second scenario, normalization was performed in LogLead, with Drain being solely responsible for parsing.

In Table \ref{tab:normalization}, the results show that using LogLEAD for normalization is 2.25, 1.64, and 1.84 times faster than using Drain normalization with Hadoop, BGL, and HDFS data, respectively. In other words, LogLead normalization takes only 44\%, 61\%, and 54\% of the time it takes to use Drain for both normalization and parsing with Hadoop, BGL, and HDFS data, respectively. This findings means that offloading normalization from Drain to LogLead is more efficient when using Drain for parsing in LogLead.

\begin{table}
\centering
\caption{Drain parsing with LogLead and Drain normalization}
\label{tab:normalization}
\begin{tabular}{l l S S}    
Data & Lines & {LogLead Norm} & {Drain Norm} \\
\hline
Hadoop & 177,592 & 6,48 & 14,58 \\
BGL & 4,747,963 & 123,14 & 202,46 \\
HDFS & 11,175,629 & 383,28 & 706,23 \\

\end{tabular}

\end{table}
In Table \ref{tab:norm-comp}, we benchmark the parsing times of three parsers and the conversion of log messages into BERT embeddings. The results indicate that Drain is the fastest parser, with LenMa and BERT embeddings being nearly 40 times slower. 

\begin{table}
\centering
\caption{Log Parsing speed comparison (LogLead normalization)}
\label{tab:norm-comp}
\begin{tabular}{l r S S S S}    
Data & \multicolumn{1}{l}{Lines} & {Drain} & {Spell} & {Lenma} & {Bert}\\
\hline
HDFS 0.5\% & 56,034 & 1.84 & 6.11 & 51.54 & 59.35 \\
HDFS 1\% & 112,363 & 3.62 & 15.10 & 88.12 & 118.38 \\ 
HDFS 2\% & 223,666 & 7.21 & 28.21 & 180.50 & 235.89 \\

\end{tabular}

\end{table}

\subsection{Comparison to related work}
The field of log feature extraction places significant emphasis on log parsing. The Logparser project lists 17 tools for log parsing \cite{zhu2019tools}, and \cite{wu2023effectiveness} compares the impact of various log representations on anomaly detection. We believe that log parsing is a crucial method for representing logs. However, given its computational intensity, log parsing should be considered only after more lightweight approaches to log representation have failed to yield the desired results.



\section{Anomaly Detector}
\subsection{Input-Output}
Anomaly Detector processes the enhanced dataframe(s) to perform anomaly detection. It can generate output in the form of dataframe(s), where each row contains prediction results. Additionally, it can provide more summarized information, such as typical machine learning performance scores including Accuracy, F1, and AUC-ROC.
\subsection{Architectural Responsibility and Functional Description}
Anomaly Detector is designed for detecting anomalies and interacts with existing machine learning or deep learning libraries, or it can employ custom-made anomaly detectors. We have integrated it with five supervised models: Decision Tree (DT), Support Vector Machine (SVM), Logistic Regression (LR), Random Forest (RF), and Extreme Gradient Boosting (XGB); four unsupervised models: Isolation Forest (IF), Local Outlier Factor (LOF), KMeans (KM), and OneClassSVM (OCSVM); and two custom models Out-of-Vocabulary Detector and Rarity Model \cite{nyyssola2023efficiency}. Adding new models is very easy,  if the model's interface adheres to the conventions of the scikit-learn library \cite{scikit-learn}.  

\subsection{Empirical Evaluation}
Table \ref{tab:ano} displays the anomaly detection results using the HDFS dataset with five different supervised algorithms and five different log message representations. The HDFS dataset is generally easy for anomaly detection, but to accentuate the differences, we made the prediction task more challenging by using only a 0.5\% of the data for training. We employed the binary weighted F1-score for performance measurement, which considers only True Positives, False Positives, and False Negatives, excluding True Negatives. This approach ensures that the large volume of normal sequences in the HDFS dataset do not skew the anomaly detection results. We used 9.5\% of the HDFS data for testing, opting for this limited testing scope due to the time demands of parsing large datasets and creating BERT embeddings. It's important to note that log parsing and embedding creation are not part of the fast Polars library. Log parsing relies on standard Python, and embedding creation involves computationally heavy deep learning processes.

The results in Table \ref{tab:ano} suggest that advanced log representations, such as those produced by Drain, LenMa, and Spell or BERT embeddings, offer limited benefits in anomaly detection. While the highest individual F1-score was achieved using a Decision Tree with Spell parsing, we observed that using `Words from events with whitespace split' resulted in the best average F1-score. 
Our findings align with those in a recent journal paper where various log representations were benchmarked in greater detail, and no clear winner emerged among the different log representations \cite{wu2023effectiveness}.

\begin{table}
\centering
\caption{Anomaly detection F1-binary trained on 0.5\% subset of HDFS data. 
}
\label{tab:ano}
\begin{tabular}{l |l l l l l |l}   
 & Words & Drain & Lenma & Spell & Bert & Average \\
\hline
DT & 0.9719 & 0.9816 & 0.9803 & 0.9828 & 0.9301 & 0.9693 \\
SVM & 0.9568 & 0.9591 & 0.9605 & 0.9559 & 0.8569 & 0.9378 \\
LR & 0.9476 & 0.8879 & 0.8900 & 0.9233 & 0.5841 & 0.8466 \\
RF & 0.9717 & 0.9749 & 0.9668 & 0.9809 & 0.9382 & 0.9665 \\
XGB & 0.9721 & 0.9482 & 0.9492 & 0.9535 & 0.9408 & 0.9528 \\
\hline
Average & 0.9640 & 0.9503 & 0.9494 & 0.9593 & 0.8500 &  \\
\end{tabular}
\end{table}

\section{Conclusion, Limitations and Future Work}
LogLead is designed to streamline benchmarking across various log datasets, representations, and anomaly detectors, thereby accelerating progress in both the science of log anomaly detection and industrial log anomaly detection projects. This goal is achieved by offering an integrated suite of numerous log loaders, enhancers, and anomaly detectors. Furthermore, using Polars ensures quicker log processing compared to traditional Pandas-based solutions.

This paper marks the inaugural release of the LogLead tool, which, so far, has only been utilized internally without feedback from the larger research community or software industry. This limitation in empirical evaluation is acknowledged, but we anticipate improvements with incoming user feedback as the tool has now been released to public.

Other limitations or future improvements include the need to add more Loaders, Enhancers, and Anomaly Detectors. While most labeled datasets from loghub \cite{zhu2023loghub} are covered, many unlabeled datasets still require inclusion. In terms of Enhancers, we have three parsers while others need integration. For example, the paper \cite{zhu2019tools} features 17 parsers. Also, there are additional embedding-based representations used in log anomaly detection studies \cite{wu2023effectiveness}. Finally, although we have integrated LogLead with several sklearn \cite{scikit-learn} models and n-gram based next event prediction \cite{mantyla2022pinpointing}, integration with notable anomaly detectors like DeepLog \cite{du2017deeplog} is pending.


\section*{Acknowledgment}
The authors would like to Shayan Hashemi for helping with LenMa and Spell parser integration. 


\bibliography{IEEEabrv,biblio}

\begin{thebibliography}{10}
\providecommand{\url}[1]{#1}
\csname url@samestyle\endcsname
\providecommand{\newblock}{\relax}
\providecommand{\bibinfo}[2]{#2}
\providecommand{\BIBentrySTDinterwordspacing}{\spaceskip=0pt\relax}
\providecommand{\BIBentryALTinterwordstretchfactor}{4}
\providecommand{\BIBentryALTinterwordspacing}{\spaceskip=\fontdimen2\font plus
\BIBentryALTinterwordstretchfactor\fontdimen3\font minus \fontdimen4\font\relax}
\providecommand{\BIBforeignlanguage}[2]{{%
\expandafter\ifx\csname l@#1\endcsname\relax
\typeout{** WARNING: IEEEtran.bst: No hyphenation pattern has been}%
\typeout{** loaded for the language `#1'. Using the pattern for}%
\typeout{** the default language instead.}%
\else
\language=\csname l@#1\endcsname
\fi
#2}}
\providecommand{\BIBdecl}{\relax}
\BIBdecl

\bibitem{lupton2021literature}
S.~Lupton, H.~Washizaki, N.~Yoshioka, and Y.~Fukazawa, ``Literature review on log anomaly detection approaches utilizing online parsing methodology,'' in \emph{2021 28th Asia-Pacific Software Engineering Conference (APSEC)}, 2021, pp. 559--563.

\bibitem{he2016experience}
S.~He, J.~Zhu, P.~He, and M.~R. Lyu, ``Experience report: System log analysis for anomaly detection,'' in \emph{2016 IEEE 27th international symposium on software reliability engineering (ISSRE)}.\hskip 1em plus 0.5em minus 0.4em\relax IEEE, 2016, pp. 207--218.

\bibitem{landauer2019framework}
M.~Landauer, F.~Skopik, M.~Wurzenberger, W.~Hotwagner, and A.~Rauber, ``A framework for cyber threat intelligence extraction from raw log data,'' in \emph{2019 IEEE International Conference on Big Data (Big Data)}.\hskip 1em plus 0.5em minus 0.4em\relax IEEE, 2019, pp. 3200--3209.

\bibitem{li2023logshrink}
X.~Li, H.~Zhang, V.-H. Le, and P.~Chen, ``Logshrink: Effective log compression by leveraging commonality and variability of log data,'' \emph{arXiv preprint arXiv:2309.09479}, 2023.

\bibitem{Vink2023Polars}
R.~Vink, ``{Polars: A DataFrame library in Rust},'' \url{https://github.com/pola-rs/polars}, 2023, accessed: 2023-11-10.

\bibitem{zhu2019tools}
J.~Zhu, S.~He, J.~Liu, P.~He, Q.~Xie, Z.~Zheng, and M.~R. Lyu, ``Tools and benchmarks for automated log parsing,'' in \emph{2019 IEEE/ACM 41st International Conference on Software Engineering: Software Engineering in Practice (ICSE-SEIP)}.\hskip 1em plus 0.5em minus 0.4em\relax IEEE, 2019, pp. 121--130.

\bibitem{Heavey2023Modern}
K.~Heavey, ``Modern polars: Performance,'' \url{https://kevinheavey.github.io/modern-polars/performance.html}, 2023, accessed: 2023-11-10.

\bibitem{xu2009detecting}
W.~Xu, L.~Huang, A.~Fox, D.~Patterson, and M.~I. Jordan, ``Detecting large-scale system problems by mining console logs,'' in \emph{Proceedings of the ACM SIGOPS 22nd symposium on Operating systems principles}, 2009, pp. 117--132.

\bibitem{lin2016log}
Q.~Lin, H.~Zhang, J.-G. Lou, Y.~Zhang, and X.~Chen, ``Log clustering based problem identification for online service systems,'' in \emph{Proceedings of the 38th International Conference on Software Engineering Companion}, 2016, pp. 102--111.

\bibitem{mantyla2022pinpointing}
M.~M{\"a}ntyl{\"a}, M.~Varela, and S.~Hashemi, ``Pinpointing anomaly events in logs from stability testing -- n-grams vs. deep-learning,'' in \emph{2022 IEEE International Conference on Software Testing, Verification and Validation Workshops (ICSTW)}.\hskip 1em plus 0.5em minus 0.4em\relax IEEE, 2022, pp. 285--292.

\bibitem{oliner2007supercomputers}
A.~Oliner and J.~Stearley, ``What supercomputers say: A study of five system logs,'' in \emph{37th annual IEEE/IFIP international conference on dependable systems and networks (DSN'07)}.\hskip 1em plus 0.5em minus 0.4em\relax IEEE, 2007, pp. 575--584.

\bibitem{yu2023nezha}
G.~Yu, P.~Chen, Y.~Li, H.~Chen, X.~Li, and Z.~Zheng, ``Nezha: Interpretable fine-grained root causes analysis for microservices on multi-modal observability data,'' in \emph{Proceedings of the 31st ACM Joint European Software Engineering Conference and Symposium on the Foundations of Software Engineering}, 2023, pp. 553--565.

\bibitem{wu2023effectiveness}
X.~Wu, H.~Li, and F.~Khomh, ``On the effectiveness of log representation for log-based anomaly detection,'' \emph{Empirical Software Engineering}, vol.~28, no.~6, p. 137, 2023.

\bibitem{he2017drain}
P.~He, J.~Zhu, Z.~Zheng, and M.~R. Lyu, ``Drain: An online log parsing approach with fixed depth tree,'' in \emph{2017 IEEE international conference on web services (ICWS)}.\hskip 1em plus 0.5em minus 0.4em\relax IEEE, 2017, pp. 33--40.

\bibitem{du2016spell}
M.~Du and F.~Li, ``Spell: Streaming parsing of system event logs,'' in \emph{2016 IEEE 16th International Conference on Data Mining (ICDM)}.\hskip 1em plus 0.5em minus 0.4em\relax IEEE, 2016, pp. 859--864.

\bibitem{shima2016length}
K.~Shima, ``Length matters: Clustering system log messages using length of words,'' \emph{arXiv preprint arXiv:1611.03213}, 2016.

\bibitem{devlin2018bert}
J.~Devlin, M.-W. Chang, K.~Lee, and K.~Toutanova, ``Bert: Pre-training of deep bidirectional transformers for language understanding,'' \emph{arXiv preprint arXiv:1810.04805}, 2018.

\bibitem{le2021log}
V.-H. Le and H.~Zhang, ``Log-based anomaly detection without log parsing,'' in \emph{2021 36th IEEE/ACM International Conference on Automated Software Engineering (ASE)}.\hskip 1em plus 0.5em minus 0.4em\relax IEEE, 2021, pp. 492--504.

\bibitem{landauer2023deep}
M.~Landauer, S.~Onder, F.~Skopik, and M.~Wurzenberger, ``Deep learning for anomaly detection in log data: A survey,'' \emph{Machine Learning with Applications}, vol.~12, p. 100470, 2023.

\bibitem{nyyssola2023efficiency}
J.~Nyyss{\"o}l{\"a} and M.~M{\"a}ntyl{\"a}, ``Efficiency of unsupervised anomaly detection methods on software logs,'' \emph{arXiv preprint arXiv:2312.01934}, 2023.

\bibitem{scikit-learn}
{Scikit-learn developers}, ``Scikit-learn: Machine learning in {P}ython,'' \url{https://scikit-learn.org}, 2023, [Online; accessed 2023-11-11].

\bibitem{zhu2023loghub}
J.~Zhu, S.~He, P.~He, J.~Liu, and M.~R. Lyu, ``Loghub: A large collection of system log datasets for ai-driven log analytics,'' in \emph{2023 IEEE 34th International Symposium on Software Reliability Engineering (ISSRE)}.\hskip 1em plus 0.5em minus 0.4em\relax IEEE Computer Society, 2023, pp. 355--366.

\bibitem{du2017deeplog}
M.~Du, F.~Li, G.~Zheng, and V.~Srikumar, ``Deeplog: Anomaly detection and diagnosis from system logs through deep learning,'' in \emph{Proceedings of the 2017 ACM SIGSAC conference on computer and communications security}, 2017, pp. 1285--1298.

\end{thebibliography}

\end{document}